\documentstyle[aps,preprint,epsfig]{revtex}

\begin{document}

\draft

\title{Anomalous diffusion associated
with nonlinear fractional derivative Fokker-Planck-like equation: Exact time-dependent solutions}
\author{Mauro Bologna$^{1}$,  Constantino Tsallis$^{1,2}$, Paolo Grigolini$^{1,3,4}$}
\address{$^{1}$Department of Physics, University of North Texas,\\
P.O. Box 311427, Denton, Texas 76203, USA }
\address{$^{2}$Centro Brasileiro de Pesquisas F\'{i}sicas, \\
Rua Xavier Sigaud 150, 22290-180 Rio de Janeiro - RJ, Brazil}
\address{$^{3}$Istituto di Biofisica CNR, Area della Ricerca di Pisa,
Via Alfieri 1, San Cataldo 56010, Ghezzano - Pisa, Italy }
\address{$^{4}$Dipartimento di Fisica dell'Universit\`{a} di Pisa and INFM,
Piazza Torricelli 2, 56127 Pisa, Italy\\
(mb0015@unt.edu, tsallis@unt.edu, tsallis@cbpf.br, grigo@unt.edu, grigo@unipi.it)}
\maketitle

\begin{abstract}
We consider the $d=1$ nonlinear Fokker-Planck-like equation with fractional derivatives 
$\frac{\partial }{\partial t}P\left( x,t \right)=D \frac{\partial^{\gamma
} }{\partial x^{\gamma
}}\left[P\left(x,t \right) \right]^{\nu }$. Exact time-dependent solutions are found for
$ \nu = \frac{2-\gamma }{ 1+ \gamma }$ ($-\infty<\gamma \leq 2$). By considering the long-distance {\it asymptotic} behavior of these solutions, a connection is established, namely $q=\frac{\gamma+3}{\gamma+1}$ ($0<\gamma \le 2$), with the solutions optimizing the nonextensive entropy characterized by index $q$ . Interestingly enough, this relation coincides with the one already known for L\'evy-like superdiffusion (i.e., $\nu=1$ and $0<\gamma \le 2$). Finally, for $(\gamma,\nu)=(2, 0)$ we obtain $q=5/3$ which differs from the value $q=2$ corresponding to the $\gamma=2$ solutions available in the literature ($\nu<1$ porous medium equation), thus exhibiting nonuniform convergence.


\end{abstract}

\newpage
A great variety of diffusive problems in nature, namely those referred to as {\it normal} diffusion, are satisfactorily described by the Fokker-Planck linear equation
\begin{equation}\label{linear}
\frac{\partial }{\partial t}P\left( {\bf x},t \right)=D\,
\nabla ^{2}P\left({\bf x},t \right)
\end{equation}
where $ P\left({\bf x},t \right)$ is the density of probability in the ${\bf x} \equiv \{x_1,x_2,...,x_d\}$ space and $D >0$ is the diffusion coefficient. Such processes are currently characterized by the fact that $ \langle {\bf x}^2 \rangle \propto t$, as shown by Einstein in his celebrated 1905 paper on Brownian motion.

More recently, several works \cite{ross} have focused on  the same type of
linear  equation but with {\it fractional} derivatives. More precisely
\begin{equation}\label{fraclinear}
\frac{\partial }{\partial t}P\left( {\bf x},t \right)=D\,
\nabla ^{\gamma}P\left({\bf x},t \right) \;\;\;\;(-\infty< \gamma \le 2)
\end{equation}
where $\nabla ^{\gamma}\equiv  \sum_{i=1}^d \frac{\partial ^{\gamma}}{\partial x_i ^{\gamma }}$ . 
Also, the {\it nonlinear}  equation with ordinary derivatives has been \cite{plastino,bukman} focused on  as well. More precisely
\begin{equation}\label{nonlinear}
\frac{\partial }{\partial t}P\left( {\bf x},t \right)=D\,
\nabla ^{2}\left[P\left({\bf x},t \right) \right]^{\nu } \;\;\;\;( \nu > -1)
\end{equation}
(no solutions are known for $\nu \le -1$ which are integrable \cite{bukman}).

These two generalized Fokker-Planck equations have been used to study anomalous L\'evy-like diffusion as well as correlated-like diffusive processes in porous media \cite{plastino,bukman,drazer,hilfer,zanette,levy,chaves,buiatti,zaslavsky,seshadri,spohn,bychuk,Strier,westgrigo,grigowest,huillet}. The present paper addresses the unification of both equations as follows:
\begin{equation}\label{tridim}
\frac{\partial }{\partial t}P\left( x,t \right)= D \nabla ^{\gamma}\left[P\left(x,t \right) \right]^{\nu}\;\;\;(-\infty<\gamma \le2)
\end{equation}
We will restrict ourselves to the $d=1$ case. More specifically, we are interested in normalized scaled solutions of the type
\begin{equation}\label{solution}
P\left(x,t \right)=\frac{1}{\phi \left(t \right)}
F\left[\frac{x}{\phi \left(t \right)} \right]
\end{equation}
Inserting this form into Eq.(\ref{tridim}) (and, without loss of generality, setting  $D=1$)
we obtain:
\begin{equation}\label{reduced}
-\frac{\dot{\phi} \left(t \right)}{\phi \left(t \right)^{2}}\left[\frac{d}{dz}
F\left(z \right)+zF\left(z \right) \right]=
\frac{1}{\phi \left(t \right)^{\nu +\gamma }}
\frac{d^{\gamma }}{dz^{\gamma }}\left[F\left(z \right)^{\nu } \right]
\end{equation}
where we have used the generic property
\begin{equation}\label{reduc}
\frac{d^{\delta  }}{dx^{\delta }}F\left( ax\right)=
a^{\delta }\frac{d^{\delta  }}{dz^{\delta }}F\left( z\right)\;\;\;\;(\delta \in {\cal R})
\end{equation}
with $ z=ax$. This basic property holds not only for the ordinary
derivative but also for all fractional operators we are aware of. By choosing
the ansatz
\begin{equation}\label{ansatz}
-\frac{\dot{\phi} \left(t \right)}{\phi \left(t \right)^{2-\nu -\gamma}}=k
\end{equation}
where $k$ is an arbitrary constant, we obtain
\begin{equation}\label{phitime}
\phi \left(t \right)=\frac{1}{\left(k_{1}t+k_{2} \right)^{\frac{1}{\nu +\gamma -1}}}       
\end{equation}

\begin{equation}\label{final}
\frac{d^{\gamma }}{dz^{\gamma }}\left[ F\left(z \right) \right]  ^{\nu }=
 k\frac{d}{dz}\left[z F\left(z \right)  \right]
\end{equation}
with $ k_{1}\equiv -\left(\gamma +\nu -1 \right)k$, $k_2$ being another arbitrary constant.
Finally making an integration we obtain
\begin{equation}\label{start}
 \frac{d^{\gamma-1 }}{dz^{\gamma -1}}\left[ F\left(z \right) \right]  ^{\nu } =
 k zF\left(z \right)+c
\end{equation}
where $c$ is another arbitrary constant.

Thus far it has not been necessary to specify  the fractional operator we refer to. Indeed, several fractional generalizations exist for the ordinary derivative, namely the Riemann-Liouville \cite{ross} (based on Laplace transform), Weyl \cite{ross} (based on Fourier transform) and Caputo \cite{caputo} (also based on Laplace transform) ones. From now on we will use the Riemann-Liouville operator, since it is for this one that it has been possible to find new exact solutions. In this case we will work with the {\it positive} $ x$ axis and, later on, we will use symmetry to extend the results to the entire real axis (we are working, in other words, with $\frac{\partial^{\gamma}}{\partial |x|^{\gamma}}$). Also, we will  use
the following generic result \cite{bologna} (see Appendix):
\begin{equation}\label{mono}
D^{\delta }_{x}\left[x^{\alpha } \left(a+bx \right)^{\beta }\right]=
a^{\delta }\frac{\Gamma \left[\alpha +1 \right]}{\Gamma \left[\alpha +1-\delta  \right]}
x^{\alpha-\delta}\left( a+bx \right)^{\beta -\delta }
\end{equation}
with $ D^{\delta }_{x}\equiv \frac{d^{\delta} }{dx^{\delta }}$ and $ \delta  \equiv \alpha+ \beta+1 $. By defining $g(x) \equiv x^{\frac{\alpha}{\nu } }\left( a+bx \right)^{\frac{\beta }{\nu }}$ and $\lambda \equiv \alpha \left(1-\frac{1}{\nu } \right)-\delta$, and rearranging the indices, Eq.(\ref{mono})  can be rewritten as follows:
\begin{equation}\label{fract2}
D^{\delta}_{x}\left[g\left(x \right) \right]^{\nu }= \frac{\Gamma
\left[\alpha +1 \right]}{\Gamma \left[\alpha +1-\delta
\right]}a^{\delta}x^{\lambda }g\left(x \right).
\end{equation}
Using this property in Eq. (\ref{start}) and, for simplicity, choosing $c=0$, we find
\begin{eqnarray}\label{parameter}
       \alpha&     =   &    \frac{\left(2-\gamma  \right)\gamma }{1-2\gamma }   \\
       \beta &    =    &   -   \frac{\gamma ^{2}-3\gamma +2}{1-2\gamma } \\
       \nu &    =    &\frac{2-\gamma }{ 1+ \gamma }.
\end{eqnarray}
These results allow us to write the solution in the form
\begin{equation}\label{solut}
      P\left(x,t \right)  =\frac{A}{\left(|k_1|t \right)^{\frac{\gamma +1}{\gamma ^{2}-\gamma +1}}}
\left[\frac{z^{\gamma \left(\gamma +1 \right)}}
{\left(1+bz \right)^{1-\gamma ^{2}}}\right]^{ \frac{1}{1-2\gamma }}
\end{equation}
\begin{equation}\label{adef}
A=\left[k\frac{\Gamma\left(\beta\right)}
{\Gamma\left(\alpha+1)\right)}\right]^{\frac{1+\gamma}{1-2 \gamma}}
\end{equation}
\begin{equation}\label{z}
z\equiv\frac{x}{\left(|k_1|t \right)^{\frac{\gamma +1}{\gamma
^{2}-\gamma +1}}}
\end{equation}
where $b$ is an arbitrary constant  (to be taken, later on, as $\pm 1$ according to the specific solutions that are studied) and where, 
without loss of generality, we have set $ k_{2}=0$ and $ a=1$. 
Indeed, the $ k_2$ constant can be incorporated into a shift of the origin of time, and $ a$ 
can be incorporated into the normalization constant $A$. We also  mention the exact solution
 
\begin{equation}\label{powerz}
F\left(z \right)\propto z^{\frac{\gamma }{\nu -1}}
\end{equation} 
that is not normalizable.
Several regions will have to be considered, 
namely
\begin{equation}\label{region}
 -\infty < \gamma<-1\,\,,\,-1< \gamma<0\,\,,\,
0 < \gamma<\frac{1}{2} \,\,,\, \frac{1}{2}< \gamma< 1\,\,,\,1< \gamma< 2
\end{equation}

We start with the region $-\infty < \gamma<-1$
for which, again without loss of generality, we can choose $b=-1$. 
The normalization condition implies

\begin{equation}\label{zoneinf}
A\int\limits_{-1}^{1}
\left[\frac{z^{\gamma \left(\gamma +1 \right)}}
{\left(1-z \right)^{1-\gamma ^{2}}}\right]^{ \frac{1}{1-2\gamma }}dz=
2A\,\frac{\Gamma \left[\frac{\gamma ^{2}-\gamma +1}{1-2\gamma }\right]
\Gamma \left[-\frac{\gamma \left(\gamma -2 \right)}{2\gamma -1} \right]}
{\Gamma \left[1-\gamma\right]}=1
\end{equation}
See Fig. 1. Also, as we can see from the limits of the different regions (\ref{region}), 
we will have to consider different particular cases, namely
$\gamma=-1\,,\gamma=0\,,\gamma=1/2\,,\gamma=1$ and $\gamma=2$.
\begin{figure}\label{mauro1}
\begin{center}
\epsfig{file=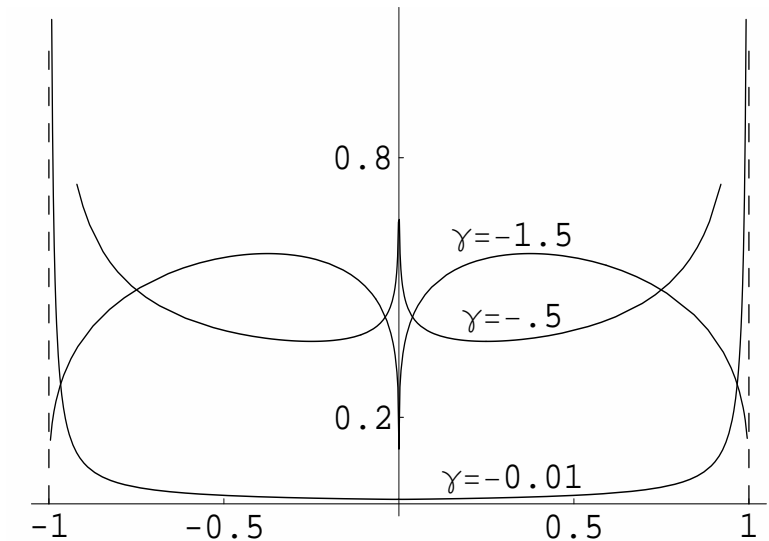,height=9cm,width=13cm,angle=0}
\caption{$(|k_1|t)^{\frac{\gamma+1}{\gamma^2-\gamma+1}}\;P\left(x,t \right)$ versus $z \equiv x/ (|k_1|t)^{\frac{\gamma+1}{\gamma^2-\gamma+1}}$ for $\gamma \le 0$; $\gamma=0^{-}$ corresponds 
to a distribution everywhere vanishing, except at the abcissa being $\pm 1$ where it diverges. For $ \gamma < -1 \left(-1< \gamma< 0  \right)$ the distribution vanishes (diverges) at $z=\pm 1$.}
\end{center}
\end{figure}

Let us start with $\gamma = -1$ and arbitrary $\nu$. 
The corresponding equation is 
\begin{equation}\label{-1}
\frac{\partial }{\partial t}P\left( x,t \right)=
\int\limits_{0}^{x}\left[P\left(y,t \right)\right]^{\nu}dy
\end{equation}
To solve it  let us go back to Eq. (\ref{final}); after  derivation with respect to $z$, 
we obtain
\begin{equation}\label{boh}
k\frac{d^2 zF(z)}{dz^2}=
\left[F\left(z\right)\right]^{\nu} 
\end{equation}
We are not going to treat this equation in detail;  we rather limit ourselves to remark that the value $\gamma=-1$ 
corresponds to $\nu \rightarrow  \pm \infty$ in the curve (16). In the region $-1 < \gamma< 0$ the probability density has a compact support, on the edges of which it diverges. See Fig. 1.

Let us now address the $0 < \gamma<  1/2$ region (where $b=1$). 
In the limit $\gamma=0$ ($\gamma=1/2$) integrability fails at infinity (at the origin); 
no such problems exist for $0 < \gamma<  1/2$. Normalization implies
\begin{equation}\label{aconst}
A=\left[ k\frac{\Gamma \left(\frac{2-3\gamma +\gamma ^2}{1-2\gamma }\right)}
{\Gamma \left(\frac{1-\gamma ^2}{1-2\gamma } \right)}\right]^{\frac{1+\gamma}{1-2 \gamma}}=
\frac{\Gamma \left(\frac{1-\gamma ^{2}}{1-2\gamma } \right)}
{2\Gamma \left(\gamma \right)
\Gamma \left( \frac{\gamma ^{2}-\gamma +1}{1-2\gamma }\right)}
\end{equation}

It is easy to show that $P\left(x,t \right)$ achieves a  maximum (see Fig. 2) at
\begin{equation}\label{max}
z=\frac{\gamma }{\left(1-2\gamma  \right)}
\end{equation}

Let us now address the $\gamma=0$ limit. It corresponds to the equation
\begin{equation}\label{zero}
\frac{\partial }{\partial t}P\left( x,t \right)=
\left[P\left(x,t \right) \right] ^{\nu}
\end{equation}
that can be resolved analytically for arbitrary $\nu $. 
To obtain this solution it is convenient to go back to Eq.(\ref{final}). It follows that
\begin{equation}\label{gam0}
F\left(z \right)=\frac{B}{z}\left[1+cz^{1-\nu } \right]^{\frac{1}{1-\nu }}
\end{equation}
Therefore, in the $ z\to \infty $ (or, equivalently $ x \to \infty $) limit, we have that $F(z) \propto 1/z$ if $\nu>1$ and $F(z)$ is a constant if $\nu<1$. The $\nu=1$ case needs specific discussion and we obtain $F(z) \propto z^{\frac{1}{k}-1}$. It is worthy reminding that the $\gamma=0$ solutions cannot be considered as distributions of probabilities  since they are not normalizable.

The $\gamma=1/2$ limiting case corresponds to the following linear equation:
\begin{equation}\label{half}
\frac{\partial }{\partial t}P\left( x,t \right)=
\frac{\partial^{\frac{1}{2}}}{\partial x^{\frac{1}{2}}}P\left(x,t \right)
\end{equation}
This equation can be solved by using the Laplace transform on both
$t$ and $x$. It can also be solved by
taking the limit $\gamma\rightarrow 1/2$. We have followed this procedure and, after
tedious though straightforward calculations, we obtain $k\sim\frac{8}{3}(1-2\gamma)$ and $A\sim [\frac{8\pi}{3}(1-2\gamma)]^{-1/2}/2$, which, replaced into Eq. (\ref{solut}), yield

\begin{equation}\label{plim}
P_{1/2}\left(x,t\right) \equiv
\lim_{\gamma\to \frac{1}{2}}P_{\gamma}\left(x,t\right)
=\frac{1}{4\sqrt{\pi}
t^{2}}\frac{\exp(-\frac{t^{2}}{4x})}{(x/t^{2})^{3/2}},
\end{equation}
which is a distribution of the Poisson type. Let us stress that the above distribution can indistinctively be obtained by solving the fractional differential equation for $\gamma=1/2$, or by taking the $\gamma >1/2$ and the $\gamma <1/2$ solutions and then considering the $\gamma \rightarrow 1/2$ limit.

Let us now focus on the $1/2 <\gamma <2$ region (where $b=-1$). The solutions strictly vanish inside an interval which contains the origin. Outside this interval, the solutions are everywhere finite if $1/2 <\gamma <1$, whereas they diverge if $1<\gamma<2$ (see Fig. 2).
\begin{figure}\label{mauro2}
\begin{center}
\epsfig{file=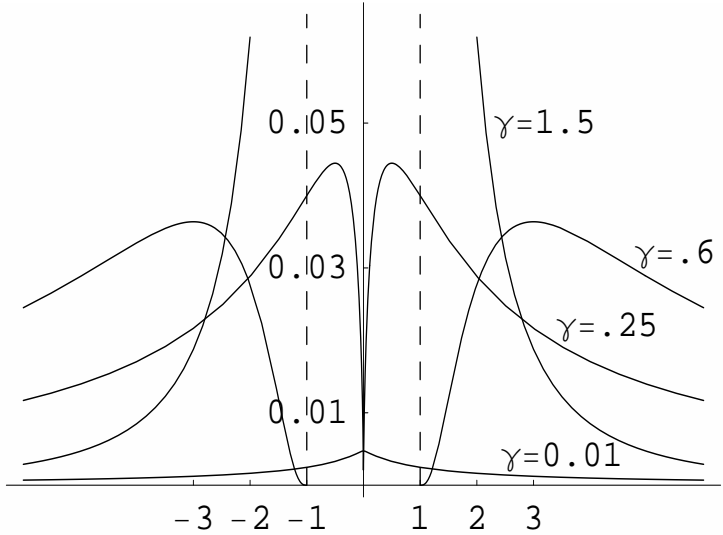,height=9cm,width=13cm,angle=0}
\caption{$(|k_1|t)^{\frac{\gamma+1}{\gamma^2-\gamma+1}}\;P\left(x,t \right)$ versus $z\equiv x/ (|k_1|t)^{\frac{\gamma+1}{\gamma^2-\gamma+1}}$ for $0 \le \gamma <2$; $\gamma=0^{+}$ 
corresponds to a distribution which vanishes everywhere. For $0<\gamma \le 1/2$ the distributions are defined in the entire real abcissa axis; for $1/2 < \gamma <1$ the distributions vanish within the $(-1,1)$ abcissa interval; for $\gamma \ge 1$, a divergence exists at the abcissa $\pm 1$ (vertical dashed asymptotes).} 
\end{center}
\end{figure}
The solutions corresponding to the $1/2 <\gamma <1$ region are as follows:
\begin{equation}\label{solutcut}
 P\left(x,t \right)  =\frac{A}{\left(kt \right)^{\frac{\gamma +1}{\gamma ^{2}-\gamma +1}}}
\left[\frac{z^{\gamma \left(\gamma +1 \right)}}
{\left(-1+z \right)^{1-\gamma ^{2}}}\right]^{ \frac{1}{1-2\gamma }}
\end{equation}
Normalization implies
\begin{equation}\label{cutoff}
2A\int\limits_{1}^{ \infty }
\left[\frac{z^{\gamma \left(\gamma +1 \right)}}
{\left(-1+z \right)^{1-\gamma ^{2}}}\right]^{ \frac{1}{1-2\gamma }}dz=
2A\, \frac{\Gamma \left[\gamma  \right]
\Gamma \left[-\frac{\gamma \left(\gamma -2 \right)}
{2\gamma -1} \right]}{\Gamma \left[\frac{\gamma \left(\gamma +1 \right)}
{2\gamma -1} \right]}=1
\end{equation}
from which $A$ is uniquely determined; finally, $k$ is obtained from $A$ by using Eq. (\ref{adef}). See Fig. 2.

Let us now focus the special case $\gamma=1$. In this case the equation becomes
\begin{equation}\label{ge1}
\frac{\partial}{\partial t}
P\left(x,t\right)=\frac{\partial}{\partial
x}P\left(x,t\right)^{\nu}\, .
\end{equation}
Its generic solution of the form indicated in Eq. (\ref{solution})) is:
\begin{equation}\label{ge1s}
[F\left(z\right)]^{\nu}=kzF\left(x,t\right)+c,
\end{equation}
which implicitly determines $F(z)$. The solution corresponding to
$c=0$ is
\begin{equation}\label{fs}
F\left(z\right)\propto z^{1/(\nu-1)}\, .
\end{equation}

In the region $1<\gamma<2$ we have the same analytic solution that we had in the region $1/2 < \gamma <1$ (i.e., Eq. (\ref{solutcut})); however, at the point $z=1$, a divergence is now present (see Fig. 2). It is clear that this solution cannot be used without appropriate asymptotic considerations for $\gamma = 2$, since for $ \gamma \to 2$,  $\nu\to 0$.

Let us now specifically focus on the possible $x \rightarrow \infty$ tail of $P(x,t)$ for arbitrary $t$. For $-\infty <\gamma <0$ the support is compact, hence there is no tail. For $0<\gamma <2$, we obtain, using either Eq. (\ref{solut}) (for $0<\gamma<1/2)$ or Eq. (\ref{solutcut}) (for $1/2<\gamma<2$), the following asymptotic behavior: 
\begin{equation}\label{asympt}
P\left(x,t \right) \sim \frac{1}{t^{\frac{\gamma +1}{\gamma ^{2}-\gamma +1}}}z^{\frac{\gamma \left(\gamma + 1 \right)+
\gamma ^{2}-1}{1-2\gamma }}
\sim \frac{t^{\frac{\gamma \left(\gamma +1
\right)}{\gamma ^{2}-\gamma +1}}}{x^{1+ \gamma }}
\end{equation}
We can easily verify that the exponent $[\gamma \left(\gamma +1
\right)] / [\gamma ^{2}-\gamma +1]$ monotonically increases from zero to $\frac{2}{\sqrt{3}}+1$ 
when $\gamma$ increases from zero to $\frac{1+\sqrt{3}}{2}$.
Also, we verify that, for $0<\gamma<1$, both $ \langle |x| \rangle$ and $\langle x^2 \rangle$ diverge; for $1<\gamma<2$, only the latter does. For $\gamma<0$, all momenta are finite since the support is compact. Finally, in all cases, $\langle x \rangle$ vanishes because of symmetry. 

Let us now address the solutions for $\nu=1$ and arbitrary $\gamma$. The equation to be 
solved becomes
\begin{equation}\label{levy}
\frac{\partial}{\partial t}
P\left(x,t\right)=\frac{\partial^{\gamma}}{\partial x^{\gamma}
}P\left(x,t\right)
\end{equation}
This equation can be solved using Laplace transform and the solutions are discussed in
\cite{metzler}. Moreover, we can see (in Fig. 3), that the point $(\gamma,\nu)=(1/2,1)$ is at the crossing of two solvable lines. Since solution (30) has been found as the limit of either one of these lines, it seems reasonable to conjecture that the same solution is found as a limit along any curve through that point.

\begin{figure}\label{mauro3}
\begin{center}
\epsfig{file=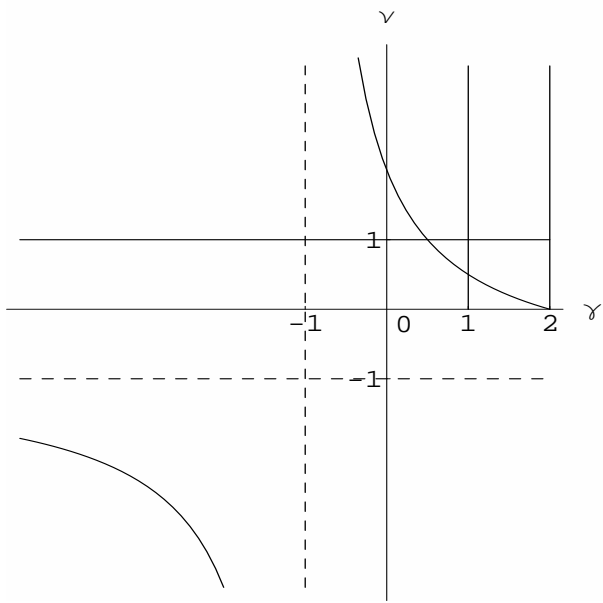,height=10cm,width=10cm,angle=0}
\caption{Curves in the $(\gamma,\nu)$ space on which exact solutions are now available: $\nu=1$ $[1,20]$, $\gamma=1$ (present work), $\gamma=2$ $[3]$, $\nu=\frac{2- \gamma}{1+\gamma}$ (present work).
The horizontal dashed line corresponds to 
the $\gamma \rightarrow -\infty$ asymptote; 
the vertical dashed line corresponds to the $\nu \rightarrow \pm \infty$ asymptote.}
\end{center}
\end{figure}
${}\\$

Let us summarize the present work. We have addressed a generic Fokker-Planck-like diffusive equation, namely the one-dimensional case of Eq. (\ref{nonlinear}), and have looked for exact scaled solutions of the type in Eq. (\ref{solution}). In the $(\gamma,\nu)$ parameter space, the solutions corresponding to $\gamma=2$ and $\nu>-1$ (porous medium equation) as well as  to $ 0<\gamma <2$  and $\nu=1$  are available in the literature, as already mentioned. We are now exhibiting exact solutions along two new lines, namely the line $\gamma=1$ and arbitrary $\nu$, and the line indicated in Eq. (16) (i.e., $\nu=(2-\gamma)/(1+\gamma)$ with $-\infty<\gamma<2$). For the latter, we observe on Eq. (36) that the spatial asymptotic behavior is characterized by the exponent $1+\gamma$, which exactly coincides with that corresponding to L\'evy superdiffusion. This is a remarkable result, since the present solutions concern a {\it nonlinear Riemann-Liouville-fractional} differential equation, !
and {\it not} the usual {\it linear Fourier-fractional} one, whose solutions are known to essentially be  L\'evy distributions. It would no doubt be interesting to know whether the same behavior is obtained {\it no matter the value of} $\nu$ ($0<\gamma <2$).

Let us finally mention a connection between the present problem and the solutions obtained from the optimization, under appropriate constraints (normalization and finite $q$-expectation value of $x^2$ in the interval $(-\infty, \infty)$), of the nonextensive entropy \cite{tsallis3,tsallis} $S_q \equiv [1-\int dx \;p(x)^q]/[q-1]$. It has been shown that these optimizing distributions {\it precisely coincide} with the solutions of the present diffusive problem for $\gamma=2$. It comes out that $q=2-\nu$ ($\nu>-1$) \cite{bukman}. Along the line indicated in Eq. (16), the exact solutions of the entropic optimization problem and the present diffusive one  {\it do not  coincide} for arbitrary value of $x$. However, comparison of the $x \rightarrow \infty$ asymptotic behaviors is possible \cite{remark}. Indeed, by identifying the behavior exhibited in (\ref{asympt}) with the behavior $1/|x|^{2/(q-1)}$ obtained \cite{plastino,bukman} for the entropic problem, we obtain 
\begin{equation}
q=\frac{\gamma+3}{\gamma+1}\;\;\;\;(0<\gamma<2)\;,
\end{equation} 
which, as commented above, {\it precisely reproduces the connection established for L\'evy distributions} \cite{levy}. By using Eq. (16), this relation can be rewritten as follows: 
\begin{equation}
q=\frac{5+2\nu}{3}\;\;\;\;(0<\nu<2)\;.
\end{equation}
(We remind that the distributions for $\nu>2$ and $\nu <-1$, i.e., $\gamma<0$, have compact support). The present nontrivial solution provides, for $(\gamma, \nu) = (2, 0)$, $q=5/3$, whereas the porous medium equation solution $q=2-\nu$ provides $q=2$. This discrepancy exhibits that the point  $(\gamma, \nu) = (2, 0)$ is a singular one, at least within the fractional derivative that we have adopted in this work.

Another point worth to be mentioned is that we have compared the present solutions with those optimizing $S_q$ defined in the interval  $(-\infty, \infty)$ and using finite $q$-expectation for $x^2$ (whereas the $q$-expectation value of $x$ vanishes). This is appropriate since, through symmetry, we have extended the solutions that we have found in the interval $(0, \infty)$ to the entire real axis. Another possibility would of course be to compare the present results in the positive real semi-axis with those optimizing the entropy $S_q$ defined in the same semi-axis and using finite $q$-expectation value of $x$. If we did so, relation (39) would be replaced by $q=(\gamma+2)/(\gamma+1)$ (see also \cite{buiatti,rajaabe}).

 Finally, last but not least, now that we have finished the presentation of the various exact solutions that emerged within the present work, it is worthy to mention that it would be very welcome the discussion of the {\it stability} of such solutions. More precisely, if we start at $t=0$ with an arbitrary distribution $P(x,0)$ and make it evolve through the present differential equation, what would be the $t \rightarrow \infty$ asymptotic distribution $P_a(x,t)$? ($a$ stands for {\it attractor}, in the space of the distributions). For instance, if the evolution is determined through the convolution product (i.e., a linear fractional-derivative Fokker-Planck equation, using a Fourier-based definition of fractional derivative), then the standard and the L\'evy-Gnedenko central limit theorems apply, and consequently the attractor $P_a(x,t)$ is either a Gaussian or a L\'evy distribution (respectively when the second cumulant is finite or infinite). If the evolution is instead det!
ermined through  a nonlinear integer-derivative Fokker-Planck equation like the one considered in \cite{bukman}, then $P_a(x,t)$ is given (as numerical verifications have shown) by the distributions which optimize the nonextensive entropy, where $x$ scales with a simple function of $t$. If the time evolution is obtained, as sometimes done, through recursive use of maps \cite{massi},  $P_a(x,t)$ can present a variety of shapes depending on the specific map which is used. Finally, in our present case (nonlinear fractional-derivative Fokker-Planck equation using a Laplace-based definition of fractional derivative), the solutions we have found might well be $P_a(x,t)$. This point, however, deserves analysis on its own.

${}\\$
\setcounter{equation}{0}
\renewcommand{\theequation}{A-\arabic{equation}}
\appendix{\textbf{APPENDIX}}

We give a short review of the property of the fractional
operator used to solve the nonlinear equation (4). The Riemann -
Liouville operator is used in many applications of fractional
calculus. The usual integral representation for this operator is:
\begin{equation}\label{riemman}
D^{\alpha}_{x}f\left(x \right)=\frac{1}{\Gamma
\left(n-\alpha  \right)} \frac{d^{n
}}{dx^{n}}\int\limits_{0}^{x}\frac{dt\; f\left(t \right)}{\left(x-t
\right)^{\alpha +1}}\;\;\;\;(n-1< \alpha < n)\;.
\end{equation}
For  the calculations in this paper we have instead used  the following equivalent form:
\begin{equation}\label{app1}
D^{\alpha}_{x}x^{\rho}= \frac{\Gamma \left(\rho+ 1 \right)}{\Gamma
\left(\rho+ 1-\alpha\right)}x^{\rho-\alpha}
\end{equation}
We have also used the generalized Leibnitz formula for this kind of fractional derivative, namely
\begin{equation}\label{app2}
D^{\alpha }_{x}\left[f\left(x \right)\, g\left(x \right) \right]=
\sum _{n=0}^{\infty }{\alpha  \choose n}D^{\alpha-n }_{x}\left[
f\left(x \right)\right]\, D^{n }_{x}\left[g\left(x \right)\right]
\end{equation}
Let us also mention that, for this operator, the following property holds under Laplace transform:
\begin{equation}\label{applapl}
{\it L}\left[D^{\alpha }_{x}f\left(x \right) \right]= s^{\alpha
}F\left(s\right)-f^{\left(\alpha-1 \right)}\left(0 \right)
\end{equation}
for $\alpha \leq 1$ and where $f^{\left(\alpha-1 \right)}\left(0
\right)$ means fractional derivative calculated in $x=0$.

Let us now show how formula (\ref{app2}) leads to  Eq. (\ref{mono}), that is used in the text. By assuming no restrictions on the parameters $ \alpha $,
$ \beta $ and $ \gamma $ and applying the generalized Leibnitz rule to the
function $ x^{\alpha }\left(a+bx \right)^{\beta }$ we obtain

\begin{equation}\label{dim}
D^{\delta }_{x}\left[x^{\alpha }\left(a+bx \right)^{\beta }\right]=
\sum ^{\infty }_{n=0}{\delta  \choose n}
D^{\delta-n }_{x}\left[x^{\alpha }\right]D^{n }_{x}\left[\left(a+bx \right)^{\beta }\right].
\end{equation}
After some algebra we obtain
\begin{equation}\label{dimm}
D^{\delta }_{x}\left[x^{\alpha }\left(a+bx \right)^{\beta }\right]=
\sum ^{\infty }_{n=0}{\delta  \choose n}
\frac{\Gamma \left(\alpha+1 \right)}{\Gamma \left(\alpha +1-\delta +n \right)}
x^{\alpha -\delta +n}
\frac{\Gamma \left(\beta +1 \right)}{\Gamma \left(\beta +1 -n\right)}
\left(-1 \right)^{n}b^{n}\left(a+bx \right)^{\beta -n}.
\end{equation}
A  closed form for this series can be achieved if $\alpha + \beta+1=\delta $. Indeed, by using the Gamma function property
$ \Gamma \left(z \right)\Gamma \left(1-z \right)=\frac{\pi }{\sin \pi z}$
we obtain 
\begin{equation}\label{finalapp}
D^{\delta }_{x}\left[x^{\alpha }\left(a+bx \right)^{\beta }\right]=
a^{\delta}\frac{\Gamma \left(\alpha +1 \right)}{\Gamma \left(\alpha +1-\delta  \right)}
x^{\alpha-\delta }\left( a+bx \right)^{\beta -\delta }\;,
\end{equation}
which essentially is Eq. (12).

For completeness, it is worthy to briefly mention here a recent variation of Riemann-Liouville operator that we have mentioned in the text, namely, the
Caputo derivative. Its definition is:
\begin{equation}\label{caputo}
^C\!D^{\alpha}_{x}f\left(x \right) \equiv \frac{1}{\Gamma
\left(m-\alpha  \right)}\int\limits_{0}^{x} \frac{dt\;f^{\left(m
\right)}\left(t \right)}{\left(x-t \right)^{\alpha +1-m}}\; .
\end{equation}
($C$ stands for Caputo). The main advantage with respect to the Riemann-Liouville operator
is that Caputo derivative of a constant is zero, which is not the case of the Riemann-Liouville one. Substantially, this kind of fractional derivative is a formal generalization of the integer derivative under Laplace transform. As disadvantage, it exhibits the fact that, whenever the derivation index is an integer number, it recovers the usual derivative {\it excepting} for an additive constant, whereas the Riemann-Liouville operator has no such disagreable property.

Finally, let us also mention the definition of Weyl fractional derivative. It is based on the properties of Fourier transform, and it is defined as follows:
\begin{equation}\label{weyldisc}
^W\!D^{\alpha }_{x}f\left(x \right)=
\sum ^{+ \infty }_{k=- \infty }\left(-\imath k \right)^{\alpha }
c_{k}\exp \left(- \imath kx \right)\;,
\end{equation}
its continuum version being
\begin{equation}\label{weylcon}
^W\!D^{\alpha }_{x}f\left(x \right)=\frac{1 }{2\pi }
\int_{- \infty }^{+ \infty } d\omega \left(-\imath \omega  \right)^{\alpha }
\hat{f}\left(\omega  \right)\exp \left(- \imath \omega x \right) \;.
\end{equation}
($W$ stands for Weyl).


\begin{references}

\bibitem{ross}K.S. Miller and B. Ross, {\it An introduction to the Fractional Differential Equations} (Wiley, New York, 1993).

\bibitem{plastino} A.R. Plastino,  A. Plastino, Physica A \textbf{222}, 347, (1995).

\bibitem{bukman} C. Tsallis and D.J. Bukman, Phys. Rev. E  \textbf{54}, R2197,
(1996).

\bibitem{drazer} G. Drazer, H.S. Wio and C. Tsallis, Phys. Rev. E  \textbf{61}, 1417 (2000).

\bibitem{hilfer}R.Hilfer and L. Anton, Phys.Rev.E, \textbf{51}, R848, (1995); L. Anton,  R. Hilfer, Phys. Rev. E, \textbf{59}, 6819, (1999); R. Hilfer, Fractals 3, 211 (1995); R. Hilfer,  in {\it Anomalous Diffusion: From Basis to Applications}, eds. R. Kutner, A. Pekalski and K. Sznajd-Weron (Springer, Berlin, 1999) page 77.

\bibitem{zanette}P.A. Alemany and D.H. Zanette, Phys. Rev. E {\bf 49}, R956 (1994); D.H. Zanette and P.A. Alemany, Phys. Rev. Lett. {\bf 75}, 366 (1995); M.O. Caceres and C.E. Budde, Phys. Rev. Lett. {\bf 77}, 2589 (1996); D.H. Zanette and P.A. Alemany, Phys. Rev. Lett. {\bf 77}, 2590 (1996).

\bibitem{levy}C. Tsallis, S.V.F Levy, A.M.C. de Souza and R. Maynard, Phys. Rev. Lett. {\bf 75}, 3589 (1995) [Erratum: {\bf 77}, 5442 (1996)]; C. Tsallis, A.M.C. de Souza and R. Maynard, in {\it L\'evy flights and related topics in Physics}, eds. M.F. Shlesinger,
G.M. Zaslavsky and U. Frisch (Springer, Berlin, 1995), p. 269; D. Prato and C. Tsallis, Phys. Rev. E {\bf 60}, 2398 (1999).

\bibitem{chaves}A.S. Chaves, Phys. Lett. A {\bf 239}, 13 (1998). 

\bibitem{buiatti}M. Buiatti, P. Grigolini and A. Montagnini,  Phys. Rev. Lett. {\bf 82}, 3383 (1999).

\bibitem{zaslavsky}A.I. Saichev, Chaos {\bf 7}, 753 (1997).

\bibitem{seshadri}V. Sechadri and B.J. West, Proc. Natl. Acad. Sci. USA {\bf79}, 4501 (1982).

\bibitem{spohn} H. Spohn \emph{J. Phys. I} \textbf{3}, 69, (1993).

\bibitem{bychuk} O.V. Bychuk and B. O'shugness, Phys. Rev. Lett. \textbf{71}, 3975, (1993).

\bibitem{Strier} D.E. Strier, D.H. Zanette, and H.S. Wio, Physica A \textbf{226}, 310, (1996).

\bibitem{westgrigo}B.J. West, P. Grigolini, R. Metzler and T.F. Nonnenmacher, Phys. Rev. E {\bf 55}, 99 (1997); B.J. West and P. Grigolini, in {\it Applications of Fractional Calculus in Physics}, ed. R. Hilfer (World Scientific, Singapore, 1998).

\bibitem{grigowest}P. Grigolini, A. Rocco and B.J. West, Phys. Rev. E {\bf 59}, 2603 (1999).

\bibitem{huillet}T. Huillet, J. Phys. A {\bf 32}, 7225 (1999).

\bibitem{caputo}M. Caputo, {\it Elasticit\`{a} e dissipazione} (Zanichelli, Bologna, 1969); M Caputo and F. Mainardi, Riv. Nuovo Cimento (Ser. II) {\bf 1}, 161 (1971).

\bibitem{bologna} M.Bologna, \emph{Derivative of Real Index} [in Italian], ed.  E.T.S. Pisa (1990).

\bibitem{metzler}R. Metzler and T.F. Nonnenmacher, Phys. Rev. E  \textbf{57}, 6409,
(1998).

\bibitem{tsallis3} C. Tsallis, J. Stat. Phys. \textbf{52}, 479, (1988).

\bibitem{tsallis} C. Tsallis, Braz. J. Phys. \textbf{29}, 1, (1999) [accessible at \newline
http://sbf.if.usp.br/WWW$_{-}$pages/Journals/BJP/Vol29/Num1/index.htm]. 

\bibitem{remark}The distributions obtained through optimization of the nonextensive entropy $S_q$, the present ones and L\'evy distributions do not generically coincide.
However, they all exhibit a power-law at long distances, hence identification of their 
respective asymptotic behaviors becomes possible.

\bibitem{rajaabe}A.K. Rajagopal and S. Abe, preprint (2000) [cond-mat/0003304].

\bibitem{massi}M. Ignaccolo and P. Grigolini, {\it A non-extensive approach to the time evolution of Lyapunov coefficients}, preprint (2000).

\end{references}
\end{document}